\documentclass[12pt]{iopart}
\usepackage{epsfig}
\newcommand{\ket}[1]{|#1\rangle}
\newcommand{\bra}[1]{\langle #1|}

\begin{document}
\title{Comment on ``Geometric phases for mixed states during cyclic 
evolutions''}
\author{Erik Sj\"{o}qvist\footnote{
Electronic address: erik.sjoqvist@kvac.uu.se}} 
\address{Department of Quantum Chemistry, Uppsala University, 
Box 518, Se-751 20 Uppsala, Sweden}
\date{\today}
\begin{abstract} 
It is shown that a recently suggested concept of mixed state 
geometric phase in cyclic evolutions [2004 {\it J. Phys. A} 
{\bf 37} 3699] is gauge dependent. 
\end{abstract}
\maketitle
Fu and Chen \cite{fu04} have proposed a concept of geometric 
phase for mixed quantal states in cyclic unitary evolution. In 
the present paper, we demonstrate that this phase is gauge 
dependent and thus not a property of the path in state space. 

Consider the unitary path 
\begin{eqnarray} 
\mathcal{C} : t\in [0,\tau] \rightarrow \rho (t) = 
U(t) \rho(0) U^{\dagger} (t) 
\end{eqnarray} 
of the density operator $\rho (0)$. The evolution is cyclic iff 
$[\rho (0),U(\tau)]=0$. In such a case, $\mathcal{C}$ is closed 
and the mixed state phase $\phi_g$ proposed in Ref. \cite{fu04} 
takes the form 
\begin{eqnarray} 
\phi_g = \sum_k w_k \phi_g^k ,  
\label{eq:fu}
\end{eqnarray} 
where $w_k$ are the time-independent eigenvalues of the density 
operator and 
\begin{eqnarray}
\phi_g^k = 
i \int_0^{\tau} \bra{\psi_k (t)} d\psi_k (t) \rangle 
\label{eq:pure}
\end{eqnarray} 
are the cyclic pure state geometric phases \cite{aharonov87} 
associated with the corresponding eigenvectors $\ket{\psi_k (t)}$ 
chosen to be periodic, i.e., $\ket{\psi_k (\tau)} = 
\ket{\psi_k (0)}, \ \forall k$. 

Now, the periodicity of the eigenvectors is preserved under 
the transformation $\ket{\psi_k (t)} \rightarrow 
\ket{\tilde{\psi}_k (t)} = e^{-i\alpha_k (t)} \ket{\psi_k (t)}$ 
if $\alpha_k (\tau) - \alpha_k (0) = 2\pi n_k$, $n_k$ integer. 
It follows from Eq. (\ref{eq:pure}) that 
\begin{eqnarray}
\phi_g^k \rightarrow \tilde{\phi}_g^k =  
i \int_0^{\tau} \bra{\tilde{\psi}_k (t)} 
d \tilde{\psi}_k (t) \rangle = \phi_g^k + 2\pi n_k ,  
\label{eq:transformedpure}
\end{eqnarray} 
i.e., $e^{i\tilde{\phi}_g^k} = e^{i\phi_g^k}$. On the other 
hand, by inserting Eq. (\ref{eq:transformedpure}) into Eq. 
(\ref{eq:pure}), we obtain 
\begin{eqnarray} 
\phi_g \rightarrow \tilde{\phi}_g = \sum_k w_k \tilde{\phi}_g^k = 
\phi_g + 2\pi \sum_k w_k n_k , 
\label{eq:transformedfu}
\end{eqnarray} 
where the additional term on the right-hand side is in general not 
an integer multiple of $2\pi$, i.e., $e^{i\tilde{\phi}_g} \neq 
e^{i\phi_g}$. 

Notice that the path $\mathcal{C}$ is invariant under 
the gauge transformation $U(t) \rightarrow U(t)V(t)$ if 
$[\rho(0),V(t)]=0, \ \forall t\in [0,\tau]$. Explicitly, 
we may take
\begin{eqnarray} 
V(t) = \sum_k e^{-i\alpha_k (t)} \ket{\psi_k (0)} \bra{\psi_k (0)}   
\end{eqnarray}
with $\alpha_k (t)$ given above. Due to its dependence upon $V(t)$ via
the integers $n_k$, it follows that $\phi_g$ is not a property of the
closed path $\mathcal{C}$. In other words, the phase concept proposed
in Ref. \cite{fu04} is not gauge invariant. 

A gauge invariant mixed state geometric phase has been proposed 
in Ref. \cite{sjoqvist00}. In the cyclic case, this phase $\gamma$ 
is determined by 
\begin{eqnarray} 
\mathcal{V} e^{i\gamma} = \sum_k w_k e^{i\phi_k^g}  
\end{eqnarray}
and is independent of $n_k$ as each phase factor in the sum of the 
right-hand side transforms invariantly. This phase may be tested 
interferometrically as an incoherent average of pure state 
interference profiles upon elimination of all the pure state 
dynamical phases in one of the arms. For cyclic evolution, 
such an analysis yields the intensity 
\begin{eqnarray}
\mathcal{I} \propto \sum_k w_k \Big| e^{i\chi} \ket{\psi_k (0)} + 
e^{i\phi_g^k} \ket{\psi_k (0)} \Big|^2 \propto 
1+\mathcal{V} \cos \big[ \chi - \gamma \big] , 
\end{eqnarray}
where $\chi$ is a variable U(1) shift applied to the other arm. 
$\gamma$ has recently been measured using nuclear magnetic resonance 
technique \cite{du03}. 

In conclusion, we have demonstrated that a quantal phase concept
recently proposed in Ref. \cite{fu04} is gauge dependent and therefore
not a property of the path in state space. Thus, this phase is neither 
experimentally testable, nor does it qualify as a geometric phase for
mixed states during cyclic evolutions.


\section*{References}
\begin{thebibliography}{99} 
\bibitem{fu04} Fu L-B and Chen J-L 
2004 {\it J. Phys. A} {\bf 37} 3699 
\bibitem{aharonov87} Aharonov Y and Anandan J 
1987 {\it Phys. Rev. Lett.} {\bf 58} 1593  
\bibitem{sjoqvist00} Sj\"{o}qvist E, Pati A K, Ekert A, 
Anandan J S, Ericsson M, Oi D K L, and Vedral V 
2000 {\it Phys. Rev. Lett.} {\bf 85} 2845 
\bibitem{du03} Du J F, Zou P, Shi M, Kwek L C, Pan J-W, 
Oh C H, Ekert A, Oi D K L, and Ericsson M 2003 
{\it Phys. Rev. Lett.} {\bf 91} 100403  
\end{thebibliography}
\end{document}